# Persistent Optical Gating of a Topological Insulator

Andrew L. Yeats,[1,2] Yu Pan,[3] Anthony Richardella,[3]
Peter J. Mintun,[1] Nitin Samarth,[3] and David D. Awschalom[1,2,*]

[1] Institute for Molecular Engineering, University of Chicago, Chicago, IL 60637, USA
[2] Department of Physics, University of California, Santa Barbara, CA 93106, USA
[3] Department of Physics, Penn. State University, University Park, PA 16802, USA
*e-mail: awsch@uchicago.edu

Topological insulators (TIs) have attracted much attention due to their spin-polarized surface and edge states, whose origin in symmetry gives them intriguing quantum-mechanical properties[1,2]. Robust control over the chemical potential of TI materials is important if these states are to become useful in new technologies, or as a venue for exotic physics. Unfortunately, chemical potential tuning is challenging in TIs in part because the fabrication of electrostatic top-gates tends to degrade material properties[3,4] and the addition of chemical dopants or adsorbates can cause unwanted disorder[5]. Here, we present an all-optical technique which allows persistent, bidirectional gating of a $(Bi,Sb)_2Te_3$ channel by optically manipulating the distribution of electric charge below its interface with an insulating $SrTiO_3$ substrate. In this fashion we optically pattern *p-n* junctions in a TI material, which we subsequently image using scanning photocurrent microscopy. The ability to dynamically write and re-write mesoscopic electronic structures in a TI may aid in the investigation of the unique properties of the topological insulating phase. The optical gating effect may be adaptable to other material systems, providing a more general mechanism for reconfigurable electronics.

Most TIs are narrow band gap semiconductors vulnerable to unintentional doping. Precise materials synthesis must be combined with a dynamic method of tuning the chemical potential to deplete bulk carriers and reach the desired level with respect to the Dirac point. Much research has focused on controlling chemical potential in TIs through electrostatic gating[6]. To this end, $SrTiO_3$ has emerged as a promising dielectric substrate due in part to its extraordinarily high permittivity at low temperature[7]. This allows a significant field effect to be applied by gating the back-side of the sample, obviating the need for a top-gate structure. However, back-gating techniques do not provide spatial control of the chemical potential in TI films, which is attractive given the particular importance of edges and interfaces in TI physics. Alternatives to electrostatic gating include the adsorption of gases on TI surfaces[8–10], electron irradiation[11], vacuum deposition of potassium[12], contact with organic molecules[13], and controlled structural deformation[14]. Exposure to synchrotron radiation during photoemission experiments has been shown to affect the chemical potential of some topological insulators[9,15], and recently to locally adjust surface band bending[16]. Persistent photodoping has also been observed in a Dirac fermion system based on a HgTe/HgCdTe quantum well[17].

We report a bidirectional optical gating effect in thin films of $(Bi,Sb)_2Te_3$ grown on $SrTiO_3$. Exposure to light with energy above the band gap of $SrTiO_3$ raises the chemical potential of the $(Bi,Sb)_2Te_3$ layer, whereas illumination with lower-energy light reduces it. We attribute this to persistent, optically-induced electrical polarization in the $SrTiO_3$ substrate caused by light passing through the semi-transparent $(Bi,Sb)_2Te_3$ layer. Figure 1 shows the evolution of the longitudinal resistance of a $(Bi,Sb)_2Te_3/SrTiO_3$ heterostructure under illumination by ultraviolet (UV) or visible light. A series of timed exposures to UV light ($\lambda$ = 375 nm, $I$ = 1 mW/m$^2$), interspersed with dark periods, was performed to demonstrate the optical gating effect and its persistence. With each exposure, the longitudinal resistance evolves consistently with a monotonic increase in the chemical potential of the $(Bi,Sb,)_2Te_3$ layer. The peak in resistance at $t = 1200$ s corresponds to the charge neutrality point of the material. At $t = 1900$ s, a series of lengthening exposures to red light ($\lambda = 635$ nm, $I = 11.8$ W/m$^2$) reverses this effect. Both effects allow continuous tuning of the chemical potential with illumination dose, though the reason for the differing kinetics is not yet clear. Low optical powers were chosen for clarity. With brighter illumination, samples can be gated in < 1 s. After initial transients, the gating effect shows minimal relaxation for 16 hours (see supplementary information).

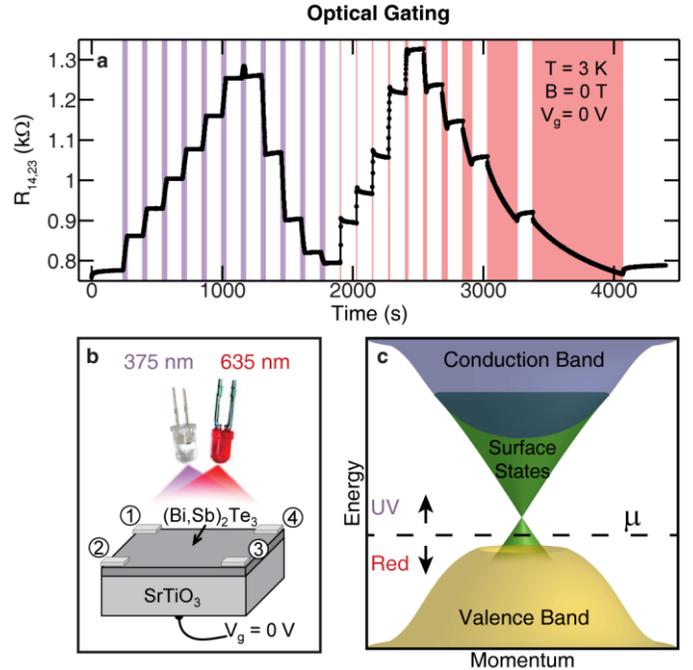

**Figure 1: Persistent optical gating of a TI channel**. **(a)** Longitudinal resistance as a function of time during UV and visible illumination. For t < 1900 s, a series of 30 s exposures to UV light (purple highlighting) followed by 120 s dark periods illustrates the persistence of the optical gating effect as the sample's chemical potential is tuned across the charge neutrality point. For t > 1900 s, a series of red light exposures (pink highlighting) reverses the effect. The exposure times were chosen for clarity given the differing kinetics of the two effects. The backside of the sample was held at 0 V for the duration of the experiment. **(b)** Schematic of measurement setup, showing Van der Pauw indices of the electrical contacts. **(c)** Schematic of the band structure of $(Bi,Sb)_2Te_3$, showing the effect of optical illumination on the chemical potential $\mu$ of the TI layer (dotted line and arrows).

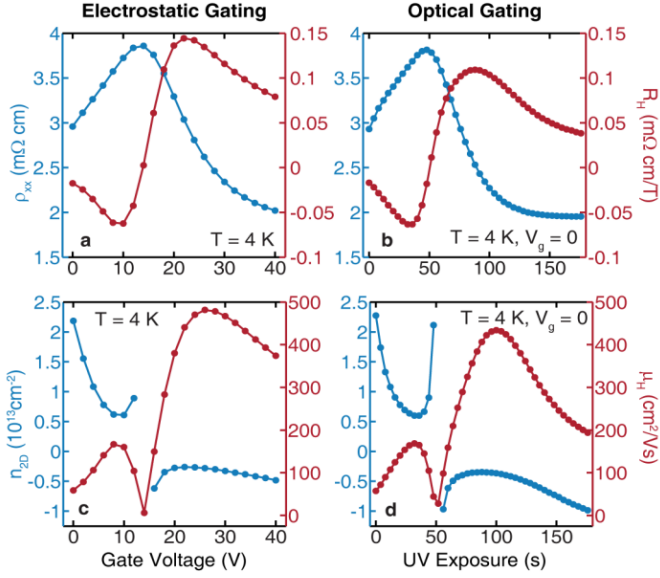

**Figure 2: Charge carrier response to electrostatic and optical gating. (a)** Longitudinal resistivity $\rho_{xx}$ (blue) and Hall coefficient $R_H$ (red) as a function of electrostatic back-gating. The peak in resistivity and change in Hall coefficient sign show the ambipolar response of the TI channel. **(b)** Resistivity and Hall coefficient after a series of consecutive timed exposures to UV light, illustrating the optical gating effect. **(c)** 2D carrier concentration $n_{2D}$ and **(d)** Hall mobility $\mu_H$ as a function of optical and electrostatic gating calculated with a one-carrier model. A data point was omitted from (c) and (d) because $n_{2D}$ diverges at $R_H = 0$ in this model. Data in all plots were collected in darkness at least 60 seconds after any illumination had ceased.

To better understand the charge carrier dynamics underlying these resistance changes, we performed Hall effect measurements as a function of both electrostatic gating and UV dose. Figure 2a shows the resistivity and Hall coefficient as a function of the potential applied to the backside of the sample. The sign change of the Hall coefficient indicates the inversion of the majority charge carrier sign. This signifies that the chemical potential has risen into the bulk band gap and past the charge neutrality point. The peak in resistivity coincides with the charge carrier inversion, suggesting that the top and bottom surfaces of the film are homogenously gated[18]. Figure 2b shows the resistivity and Hall coefficient as a function of UV exposure with the backside of the sample held at 0 V. The sample was first initialized by exposure to red light, and then subjected to a series of timed exposures to UV light. The longitudinal and Hall resistances were measured at magnetic fields between ±1 T during the dark periods between exposures. A similar response is seen for both electrostatic and optical gating. The Hall response was linear below 1 T for both types of gating, allowing interpretation with a simple one-carrier model. Figures 2c,d show the calculated 2D carrier concentration and Hall mobility. Carrier concentrations below $5\times10^{12}$ cm$^{-2}$ were observed for both types of gating, approaching reported values for the surface states' contribution in this material[19]. Samples were illuminated from above, so shadowing from the Van der Pauw contacts and surface imperfections may affect the homogeneity of optical gating in this sample, explaining the 10% reduction of peak mobility.

Further insight into the nature of the charge carriers can be gained through magnetoconductance experiments. The $\pi$ Berry phase of TI surface states requires that backscattering be suppressed in the presence of time-reversal symmetry[1]. This gives rise to a low-field correction to TIs' magnetoconductance known as weak anti-localization (WAL)[7]. Gate-tunable WAL has been reported in TIs, and its tunability attributed to surface-bulk scattering[20]. Figure 3 shows the magnetoconductance of a $(Bi,Sb)_2Te_3/SrTiO_3$ heterostructure as a function of either optical or electrostatic gating. A zero-field cusp develops as the sample is subjected to positive gate voltage or after consecutive exposures to UV light. This is consistent with the enhancement of WAL as the chemical potential of the TI channel rises above the top of the valence band into the bulk band gap and surface-bulk scattering is reduced. These data demonstrate the relevance of the optical gating effect for the study of coherent carrier effects in TI materials.

$SrTiO_3$ is a host to several optoelectronic effects which could complicate these data[21,22]. To rule out the influence of substrate conductivity effects, identically-annealed bare $SrTiO_3$ substrates were measured at 3 K and at room temperature. Two-terminal resistance measurements exceeded 10 G$\Omega$ regardless of the sample's illumination history to UV or visible light, indicating that parallel conduction through the substrate is unlikely to affect our data.

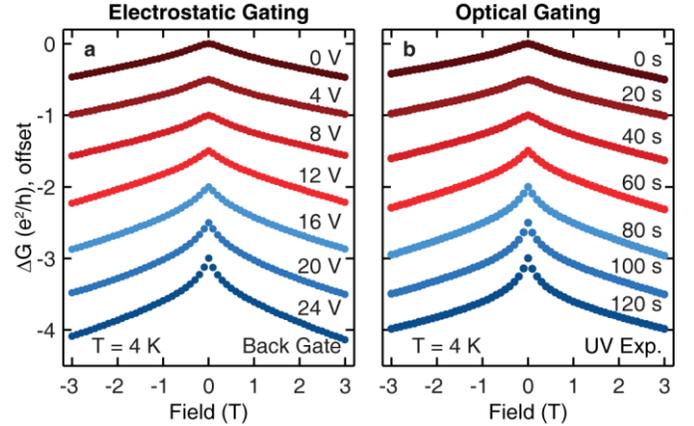

**Figure 3: Optical and electrostatic tuning of weak anti-localization.** Magnetoconductance $\Delta G$ of a $(Bi,Sb)_2Te_3/SrTiO_3$ heterostructure as a function of **(a)** electrostatic and **(b)** optical gating. A zero-field cusp develops when the sample is subjected to a positive back-gate voltage or exposed to UV light. This is consistent with enhancement of WAL as the chemical potential rises above the top of the valence band and surface-bulk scattering is reduced. Qualitatively similar behavior is seen with both gating techniques. Data in all plots were collected in darkness at least 60 s after any illumination had ceased. Data are offset for clarity.

Figure 4a shows the spectral dependence of the optical gating effect. The longitudinal resistance of the sample was measured before and after timed exposures to different energies of light from a Xe lamp and monochromator. Red and UV light were used before each exposure to initialize the sample to a regime where resistance changes would map monotonically to chemical potential (see inset). The sign of the optical gating effect inverts at the band gap energy of $SrTiO_3$, indicating that the gating effect originates in the $SrTiO_3$ substrate. Moreover, measurements of $(Bi,Sb)_2Te_3$ films grown on InP did not show a bidirectional optical response, further confirming this interpretation.

We propose a defect-mediated mechanism in $SrTiO_3$ to account for the bidirectional optical gating behavior of our samples. Persistent photorefractive effects in $SrTiO_3$ and other perovskites have been attributed to the optical manipulation of charged defect populations[23,24]. This suggests optical modulation of defect charge states as a means to produce large persistent electric fields in these materials. In our experiments, light with energy above the $SrTiO_3$ bulk band gap penetrates only a short distance past the semi-transparent TI layer and into the substrate. Carriers photoexcited from defects near the interface may recombine with states farther into the $SrTiO_3$, where the illumination is weaker, and they will be less likely to be re-excited. This produces a charge asymmetry in the defect population below the interface, growing until all illuminated defects are depopulated, or the induced electric field balances the diffusive pressure set up by the optical gradient. This field may be responsible for the gating effect on the $(Bi,Sb)_2Te_3$ layer. Absent illumination, charges will remain bound in their traps and the gating effect will persist. Alternatively, light with energy below the $SrTiO_3$ bulk band gap reaches defects throughout the substrate volume, allowing isotropic redistribution of charges and the consequent relaxation of the asymmetry. This model qualitatively explains the spectral dependence and persistence of the effect we observe, and is discussed further in the supplementary information.

Part of the interest in TIs is due to the supposition that the topological nature of their bands may allow unusual quantum effects to become relevant at ambient temperatures. Figure 4b shows the temperature dependence of the optical gating effect. The strength of the effect weakens but does not vanish as temperature increases to 295 K. This is roughly consistent with the temperature dependence of the dielectric constant of $SrTiO_3$[25]. The 295 K trace is monotonic because optical gating is not strong enough at room temperature to tune this sample through its charge neutrality point. However, optimization of $SrTiO_3$ materials properties might amplify the optical gating effect, increasing its relevance for room-temperature applications.

Persistent optical effects are amenable to spatial patterning. By selectively exposing different areas of our samples to measured doses of UV or visible light, we can create arbitrary chemical potential landscapes in a TI channel, which persist for hours after illumination. The bidirectional nature of this effect allows these patterns to be dynamically modified and re-written *in situ*, which may be useful for rapid characterization of TI electronic structures. These chemical potential landscapes may be detected with scanning photocurrent microscopy. When electron-hole pairs are photoexcited in a region with a strong chemical potential gradient, they will tend to drift apart. This produces a net photocurrent whose longitudinal component can be detected between the two current leads of a Hall bar. By rastering a laser spot over the surface of a sample while continuously monitoring the zero-bias photocurrent, the gradient of the chemical potential landscape can be imaged.

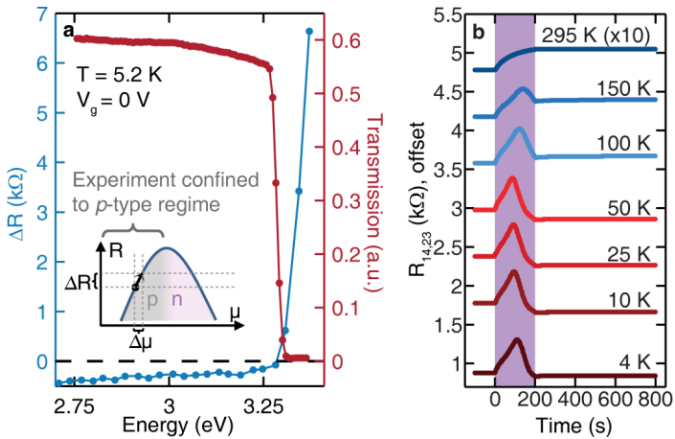

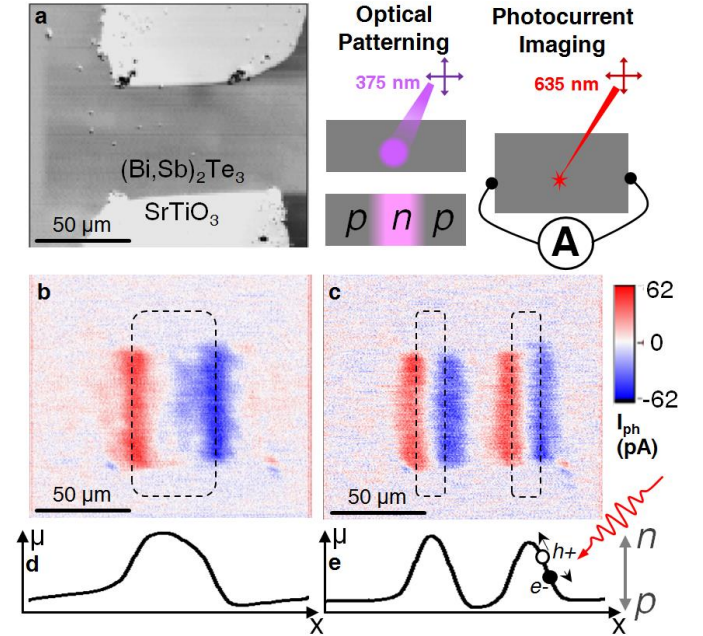

**Figure 4: Spectral and temperature dependence.**
**(a)** Spectral dependence of the optical gating effect. The relative change in longitudinal resistance Δ*R* (blue) is shown due to identically timed exposures to different energies of light. Before each exposure, red and UV light were used to reset the chemical potential to a similar starting position in the *p*-type regime such that Δ*R* maps roughly to chemical potential shift Δ*μ* (see inset). The transmission spectrum (red) of an identically-annealed $SrTiO_3$ substrate is shown for comparison. Both measurements were conducted at 5.2 K. **(b)** Temperature dependence of the optical gating effect, showing its persistence to room temperature. Longitudinal resistance is plotted as a function of time at various temperatures before, during (purple highlighting) and after UV illumination. Traces are offset for clarity.
The 295 K trace is multiplied by 10.

**Figure 5: Writing and imaging *p-n* junctions in a TI.**
**(a)** Scanning reflectance image of a $(Bi,Sb)_2Te_3$ channel. **(b,c)** Photocurrent images of the same region showing the longitudinal component of chemical potential gradients patterned with the optical gating technique. In each image the field of view was first initialized *p*-type by exposure to red light from a HeNe laser. Rectangular areas (dotted lines) were exposed to UV light before imaging, locally gating these regions *n*-type. UV exposure and photocurrent imaging were repeated and the images averaged to reduce noise. **(d,e)** Schematics generated by numerical integration of (b,c), depicting the chemical potential as a function of lateral position on the channel. The temperature was 5 K.

Photocurrent images of two chemical potential landscapes patterned with the optical gating technique are shown in Figure 5b,c. For each image the field of view was first initialized *p*-type by exposure to red light. Rectangular *n*-type regions were defined by projecting UV light onto the sample surface. Numerical integration was used to extract the induced chemical potential changes along the channel (Figure 5d,e). With this technique, arbitrary configurations of *p*- and *n*-type regions can be patterned with a resolution of < 20 μm. Exposure to the imaging beam degrades the pattern by *p*-gating the material. However, patterns imaged hours after writing show little evidence of degradation.

We have demonstrated a persistent, bidirectional optical effect which can be used to dynamically tune the chemical potential of a TI/SrTiO$_3$ heterostructure. This enables us to write and re-write arbitrary chemical potential landscapes in a TI channel without the need for additional materials or lithography. This technique may facilitate the study of proposed TI phenomena such as topological edge states at *p-n* junctions[26,27], Klein tunneling[28], spin filtering[29], bound states outside the Dirac cone[30] and Mach-Zehnder interferometry[31]. Through a procedure of incremental writing and erasing, closed regions gated to different potentials could be moved adiabatically, which may have relevance for proposed methods of quantum computation[1,2]. The optical gating effect may also be adaptable to help study other material systems grown or deposited on SrTiO$_3$ such as heterostructures with LaAlO$_3$[32], graphene[33], and transition-metal dichalcogenides[34,35]. Integration with traditional semiconductor materials might allow room temperature optical patterning and dynamic modification of electronic structures for use as reconfigurable logic or memory devices.

**Methods**

Samples consist of six quintuple-layers (QL) of (Bi,Sb)$_2$Te$_3$ grown by molecular beam epitaxy (MBE) on 5×5×0.5 mm (111)-oriented SrTiO$_3$ substrates. A Bi:Sb ratio of approximately 1:1 was chosen to position the Fermi level close to the bulk band gap[19]. Before gating, all samples showed *p*-type conductivity with a 2D carrier concentration of order $10^{13}$ cm$^{-2}$. The (Bi,Sb)$_2$Te$_3$ film thickness was chosen to avoid hybridization of the top and bottom surface states, while remaining thin enough to effectively gate both surfaces[36]. Before growth, the SrTiO$_3$ substrates were annealed in oxygen for two hours at 875 – 925 °C to improve the quality of the surface. A control sample was grown on (111)-oriented InP to rule out optical gating effects intrinsic to the (Bi,Sb)$_2$Te$_3$ material. Identically annealed bare SrTiO$_3$ substrates were also measured to control for substrate photoconductivity.

Samples grown on SrTiO$_3$ were measured either in the Van der Pauw geometry or after mechanically scratching away the growth layer to form a Hall bar. A Hall bar was patterned on the InP sample with standard photolithography techniques. An AC resistance bridge was used to measure Van der Pauw geometry samples. Standard lock-in techniques were used to measure the Hall bar geometry samples. The excitation current was 30 – 300 nA at 14 – 19 Hz. Electrical contacts to all samples were made with indium and were ohmic. Measurements were conducted in a magneto-optical cryostat. Illumination was provided by commercially-available single-color LEDs or a Xe lamp coupled to a monochromator. Photocurrent images were acquired by rastering a focused HeNe laser spot ($\lambda$ = 633 nm, $P$ = 45 μW, $d \approx 1$ μm) over the surface of the sample while monitoring the induced zero-bias photocurrent.

**Acknowledgments**

This work is supported by the ONR (Grant Nos. N00014-12-1-0116 and -0117), the AFOSR MURI (Grant No. FA9550-15-1-0029) , the ARO (Grant No. W911NF-12-1-0461), and the NSF MRSEC (Grant No. NSF-DMR-1420709). We thank Abram Falk, Bob Buckley, David Christle, Chetan Nayak and Yu-Sheng Chen for useful discussions. We thank Justin Jureller and Daniel Silevitch for assistance with the Hall measurements. We acknowledge use of the NSF National Nanofabrication Users Network Facility at Penn State.

**Author Information**

**Contributions:** Y.P. and A.R. grew, fabricated, and characterized the samples. A.L.Y., P.J.M., and Y.P. performed the optical experiments. D.D.A and N.S. supervised the efforts. All of the authors contributed to analysis of the data, discussions, and the production of the manuscript.

**Competing Financial Interests:** The authors declare no competing financial interests.

**Corresponding Author:** Correspondence should be addressed to David D. Awschalom: awsch@uchicago.edu